\def\nk{n_{\rm b}}

\def\Pb{P_{\rm b}}

\def\rfr#1{Equation\,(\ref{#1})}
\def\rfrs#1#2{Equations\,(\ref{#1})-(\ref{#2})}

\def\Rfrs#1#2{Equations\,(\ref{#1})-(\ref{#2})}

\def\virg#1{``#1"}

\def\eqi{\begin{equation}}
\def\eqf{\end{equation}}
\def\eqia{\begin{eqnarray}}
\def\eqfa{\end{eqnarray}}

\def\rp#1#2{{#1\over#2}}
\def\lb#1{\label{#1}}

\def\bds#1{\boldsymbol{#1}}

\def\ton#1{\left(#1\right)}
\def\qua#1{\left[#1\right]}
\def\grf#1{\left\{#1\right\}}

\documentclass[onecolumn]{aastex}
\usepackage{morefloats}
\usepackage[title]{appendix}
\usepackage{hyperref}
\usepackage{booktabs}
\usepackage[table,xcdraw]{xcolor}
\usepackage{multirow}
\usepackage{rotating,tabularx}
\usepackage{float}
\usepackage{enumerate}
\usepackage{rotating}
\usepackage[polutonikogreek,english]{babel}
\usepackage{amsmath,starfont,textgreek,w-greek,wasysym}
\usepackage[flushleft]{threeparttable}
\usepackage{amsthm}
\usepackage{amscd,lineno}
\usepackage{amssymb,dsfont}
\usepackage{graphicx,epsfig}
\usepackage{txfonts}
\usepackage{xr-hyper}

\RequirePackage{color}

\newcommand{\emaila}{lorenzo.iorio@libero.it}

\allowdisplaybreaks[1]

\begin{document}

\title{Is it possible to measure the Lense-Thirring orbital shifts of the short-period S-star S4716 orbiting Sgr A$^\ast$?}

\shortauthors{L. Iorio}

\author{Lorenzo Iorio\altaffilmark{1} }
\affil{Ministero dell'Istruzione e del Merito
(M.I.U.R.)
\\ Viale Unit\`{a} di Italia 68, I-70125, Bari (BA),
Italy}

\email{\emaila}

\begin{abstract}
The \textcolor{black}{maximal} values of the general relativistic Lense-Thirring (LT) orbital shifts $\Delta I^\mathrm{LT},\,\Delta\Omega^\mathrm{LT}$ and $\Delta\omega^\mathrm{LT}$ of the inclination $I$, the longitude of the ascending node $\Omega$ and the perinigricon $\omega$ of the recently discovered star S4716, which has the shortest orbital period $\ton{\Pb=4.02\,\mathrm{yr}}$ of all the S-stars that orbit the supermassive black hole (SMBH) in Sgr A$^\ast$, are of the order of $\simeq 5-16$ arcseconds per revolution $\ton{^{\prime\prime}\,\mathrm{rev}^{-1}}$. Given the current error $\sigma_\omega = 0.02^\circ$ in determining $\omega$, which is the most accurate orbital parameter of S4716 among all those affected by the SMBH's gravitomagnetic field through its angular momentum ${\bds J}_\bullet$,  about 48 yr would be needed to reduce $\sigma_\omega$ to $\simeq 10\%$ of the cumulative LT perinigricon shift over the same time span. Measuring  $\Delta I^\mathrm{LT}$ and $\Delta\Omega^\mathrm{LT}$ to the same level of accuracy would take even much longer.
Instead, after just 16 yr, a per cent measurement of the larger gravitoelectric (GE) Schwarzschild-like perinigricon shift $\Delta\omega^\mathrm{GE}$, which depends only on the SMBH's mass $M_\bullet$, would be possible. On the other hand, the uncertainties in the physical and orbital parameters entering $\Delta\omega^\mathrm{GE}$ would cause a huge systematic bias of $\Delta\omega^\mathrm{LT}$ itself. The SMBH's quadrupole mass moment $Q_2^\bullet$ induces orbital shifts as little as $\simeq 0.01-0.05\,^{\prime\prime}\,\mathrm{rev}^{-1}$.
\end{abstract}

keywords{
Kerr black holes (886); Orbital elements (1177); General relativity (641);
}
\section{Introduction}
The recently discovered star S4716  \citep{2022ApJ...933...49P} orbits the supermassive black hole\footnote{The existence of such an object in the GC was postulated for the first time by \citet{1971MNRAS.152..461L}.}  (SMBH)  Sgr A$^\star$ at the Galactic center (GC) \citep{2008ApJ...689.1044G,2010RvMP...82.3121G} in just $4.02\,\mathrm{yr}$ along an elliptical orbit whose eccentricity is $e = 0.756$. To date, it is the star with the shortest orbital period $\Pb$ among the members of the stellar S cluster around Sgr A$^\star$ \citep{2020ApJ...896..100A}; other recently discovered short period S-stars are S62 ($\Pb = 9.9\,\mathrm{yr}$) and  S4711 ($\Pb = 7.6\,\mathrm{yr}$) \citep{2020ApJ...899...50P}.

Given its closeness to its parent SMBH, S4716 may be, at least in principle, an ideal candidate to measure its gravitomagnetic Lense-Thirring (LT) orbital precessions induced by the SMBH's angular momentum
\eqi
J_\bullet = \chi_g\,\rp{M_\bullet^2\,G}{c}. \lb{SBH}
\eqf
In \rfr{SBH}, $G$ is the Newtonian constant of gravitation, $c$ is the speed of light in vacuum,  $M_\bullet$ is the hole's mass.
In the case of a rotating Kerr BH \citep{1963PhRvL..11..237K}, the theoretical upper limit $\left|\chi_g\right|\leq 1$ holds for the dimensionless spin parameter $\chi_g$ entering \rfr{SBH}. If $\left|\chi_g\right| > 1$, a naked singularity without a horizon would occur, along with the possibility of causality violations due to closed timelike curves \citep{1983mtbh.book.....C}. Incidentally, it is just the case of recalling that, although not yet proven, the cosmic censorship conjecture \citep{2002GReGr..34.1141P} states that naked singularities cannot be formed via the gravitational collapse of a body. Moreover, according to the so-called \virg{no-hair} theorems \citep{1967PhRv..164.1776I,1968CMaPh...8..245I,1971PhRvL..26..331C}, a BH without electric charge is completely characterized by its mass $M_\bullet$ and angular momentum $J_\bullet$. Then, all the multipole moments of its external spacetime are functions of just $M_\bullet$ and $J_\bullet$.
For other studies involving the S-stars and various consequences of the gravitomagnetic field of Sgr A$^\ast$, see \citet{1998AcA....48..653J,2003ApJ...590L..33L,2007CQGra..24.1775K,2008IAUS..248..466G,2008ApJ...674L..25W,2009ApJ...703.1743P,
2010ApJ...720.1303A,2010PhRvD..81f2002M,2011MNRAS.411..453I,2013degn.book.....M,2014RAA....14.1415H,2015ApJ...809..127Z,2016ApJ...818..121P,
2016ApJ...827..114Y,2017ApJ...834..198Z,2018MNRAS.476.3600W,2019GReGr..51..137P,2020arXiv200811734F,2022PhRvD.106l3023A,2022ApJ...932L..17F}.

In this paper, the perspectives of measuring the gravitomagnetic spin-induced orbital precessions of S4716 are investigated.
As far as the  gravitoelectric (GE), Schwarzschld-like perinigricon precession is concerned, it was recently measured with the S2 star to a $15\%$ accuracy level \citep{2020A&A...636L...5G}. Such an effect is \textcolor{black}{analogous to} the formerly anomalous perihelion precession of Mercury of $42.98\,\mathrm{arcseconds\,per\,century}\,\ton{^{\prime\prime}\,\mathrm{cty}^{-1}}$ \citep{1986Natur.320...39N}, successfully explained by the Einstein's general relativity at its inception \citep{Ein15}. It is due to the static part of the spacetime metric, and depends only on the hole's mass.

The paper is organized as follows. In Section\,\ref{sec2}, the gravitomagnetic LT orbital precessions for an arbitrary orientation of the primary's spin axis are reviewed. Section\,\ref{sec3} deals with the LT shifts per revolution of S4716 and the time required to measure them with a chosen accuracy level.
The same analysis is performed for the GE shift per orbit of S4716 in Section\,\ref{sec4}. The impact of the hole's quadrupole mass moment on the orbit of S4716 is dealt with in Section\,\ref{sec5}. Section\,\ref{sec6} summarizes the findings of the paper and offers conclusions.
\section{The Lense-Thirring orbital precessions}\lb{sec2}
It turns out that, for an arbitrary orientation of the primary's spin axis in space, the LT precessions of the orbital inclination $I$, the longitude of the ascending node $\Omega$, and the argument of perinigricon $\omega$ of an orbiting test particle are \citep{1975PhRvD..12..329B,1988NCimB.101..127D,1992PhRvD..45.1840D,1999ApJ...514..388W,2007CQGra..24.1775K,2008ApJ...674L..25W,2017EPJC...77..439I}
\begin{align}
\dot I^\mathrm{LT} \lb{Irate}& = \rp{2\,G\,J\,\bds{\hat{J}}\bds\cdot\bds{\hat{l}}}{c^2\,a^3\,\ton{1-e^2}^{3/2}}, \\ \nonumber\\
\dot \Omega^\mathrm{LT} \lb{Orate}& = \rp{2\,G\,J\,\csc I\,\bds{\hat{J}}\bds\cdot\bds{\hat{m}}}{c^2\,a^3\,\ton{1-e^2}^{3/2}}, \\ \nonumber\\
\dot \omega^\mathrm{LT} \lb{orate}& = -\rp{2\,G\,J\,\bds{\hat{J}}\bds\cdot\ton{2\,\bds{\hat{h}} + \cot I\bds{\hat{m}}}}{c^2\,a^3\,\ton{1-e^2}^{3/2}},
\end{align}
where $a$ is the orbit's semimajor axis.
In \rfrs{Irate}{orate}, $\bds{\hat{l}}=\grf{\cos\Omega,\,\sin\Omega,\,0}$ is the unit vector directed in the reference $\grf{x,\,y}$ plane along the \textcolor{black}{line of nodes} toward the ascending node, $\bds{\hat{m}}=\grf{-\cos I\,\sin\Omega,\,\cos I\,\cos\Omega,\,\sin I}$ is the unit vector directed in the orbital plane transversely to $\bds{\hat{l}}$, and $\bds{\hat{h}}=\grf{\sin I\,\sin\Omega,\,-\sin I\,\cos\Omega,\,\cos I}$ is the unit vector directed along the orbital angular momentum perpendicularly to the orbital plane in such a way that $\bds{\hat{l}},\,\bds{\hat{m}},\bds{\hat{h}}$ are a right-handed triad of unit vectors. \textcolor{black}{It should be remarked that \rfrs{Irate}{orate} hold in an arbitrary coordinate system, and $\bds{\hat{J}}$ can assume any orientation with respect to it. Such a choice is suitable for the case at hand since the spin axis of Sgr A$^\ast$ is poorly constrained \citep{2007A&A...473..707M}. }
By parameterizing $\bds{\hat{J}}$ as
\begin{align}
{\hat{J}}_x \lb{Jx}& = \sin i\,\cos\varepsilon, \\ \nonumber \\
{\hat{J}}_y \lb{Jy}& = \sin i\,\sin\varepsilon, \\ \nonumber \\
{\hat{J}}_z \lb{Jz}& = \cos i,
\end{align}
\rfrs{Irate}{orate} can be rewritten as
\begin{align}
\dot I^\mathrm{LT} \lb{dIdt}& =\rp{2\,G\,J\,\sin i\,\cos\zeta}{c^2\,a^3\,\ton{1-e^2}^{3/2}}, \\ \nonumber \\
\dot \Omega^\mathrm{LT} \lb{dOdt}& =\rp{2\,G\,J\,\ton{\cos i +\cot I\,\sin i\,\sin\zeta}}{c^2\,a^3\,\ton{1-e^2}^{3/2}}, \\ \nonumber \\
\dot\omega^\mathrm{LT} \lb{dodt}& = \rp{G\,J\,\qua{-6\,\cos I\,\cos i +\ton{1 -3 \cos 2 I}\,\csc I\,\sin i\,\sin\zeta}}{c^2\,a^3\,\ton{1-e^2}^{3/2}},
\end{align}
where $\zeta\doteq\varepsilon-\Omega$.
\Rfrs{dIdt}{dodt} show that the LT precessions depend explicitly on the absolute orientation of $\bds{J}$ and of the test particle's orbital plane in space through $i$ and $I$, respectively, and on the relative orientation of $\bds{J}$ and the test particle's orbit through the angle $\zeta$. Thus, they can even vanish for certain combinations of $i,\,I,$ and $\zeta$.
\section{The Lense-Thirring orbital precessions of S4716 and its measurability}
\begin{figure}[!htb]
\begin{center}
\centerline{
\vbox{
\begin{tabular*}{\textwidth}{c@{\extracolsep{\fill}}c}
\epsfysize= 6.0 cm\epsfbox{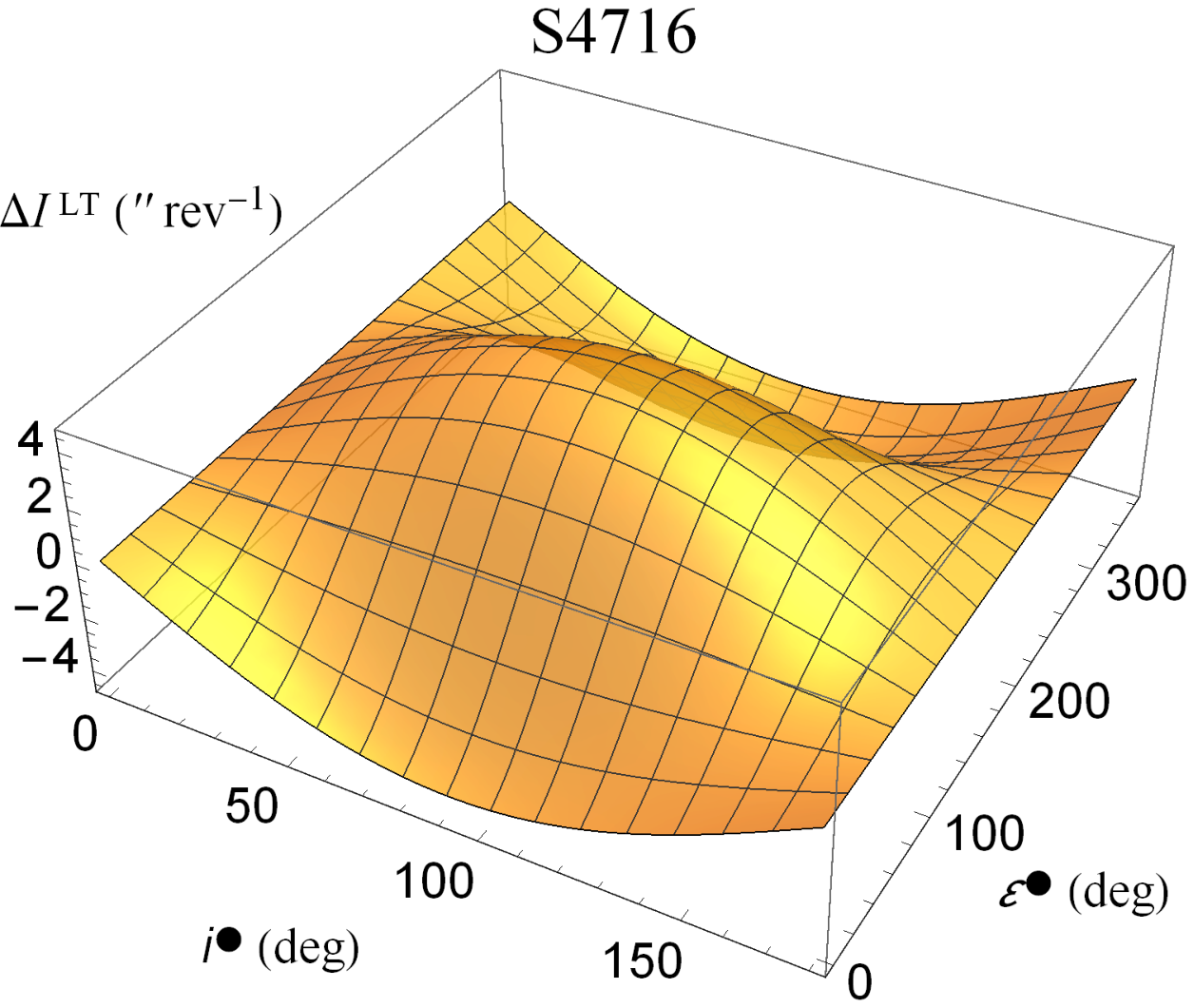}\\
\epsfysize= 6.0 cm\epsfbox{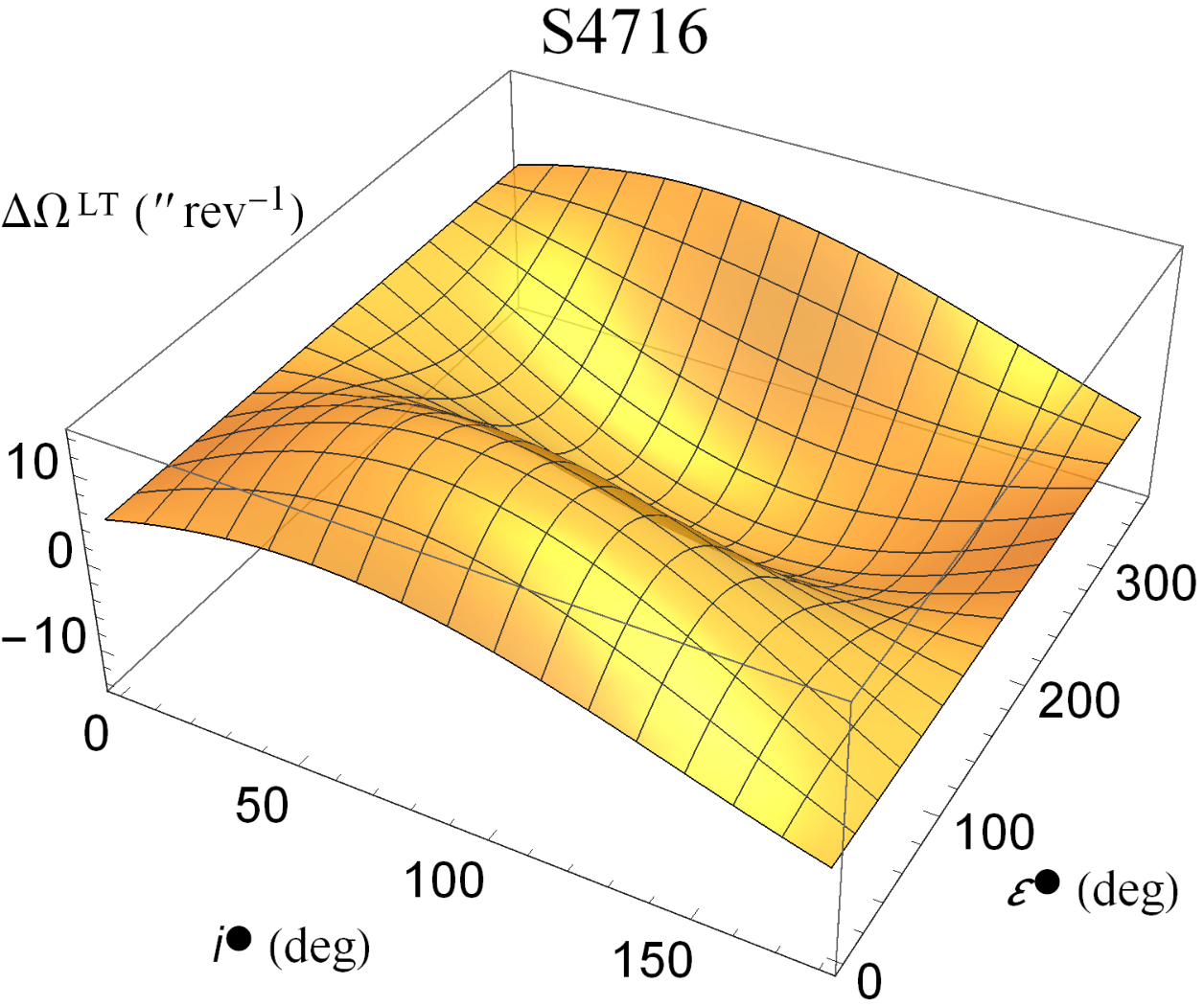}\\
\epsfysize= 6.0 cm\epsfbox{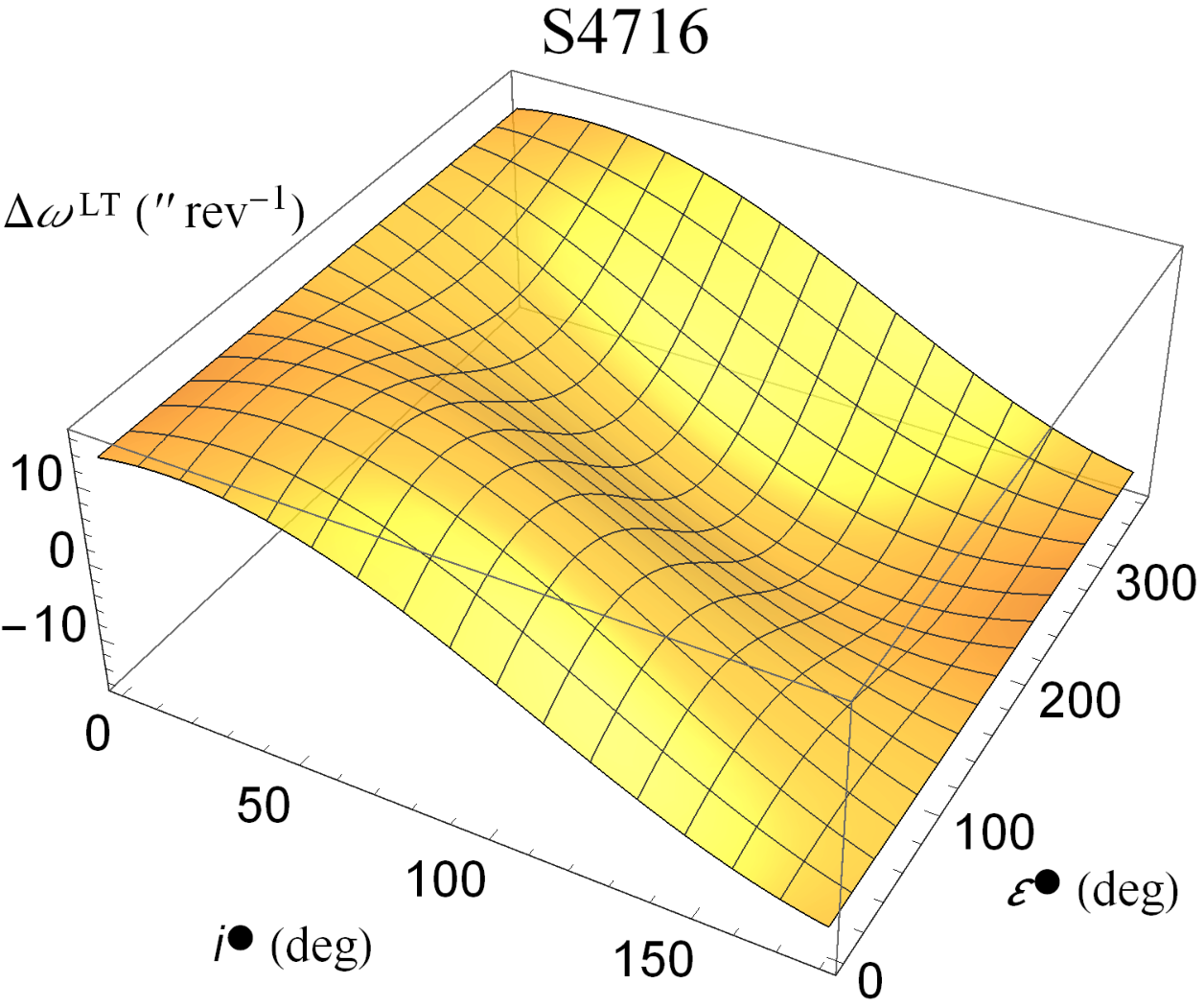}\\
\end{tabular*}
}
}
\caption{LT shifts per orbit, in arcseconds per revolution $\ton{^{\prime\prime}\,\mathrm{rev}^{-1}}$, of S4716 according to \rfrs{dIdt}{dOdt} plotted as functions of $i^\bullet,\,\varepsilon^\bullet$. For  $J_\bullet=\chi_\mathrm{g}\,M_\bullet^2\,G/c$, the values  $\chi_\mathrm{g}=0.5,\,M_\bullet=4.1\times 10^6\,\mathrm{M}_\odot$ \citep{2020ApJ...899...50P,2022ApJ...933...49P} were adopted.}\label{figura1}
\end{center}
\end{figure}

\lb{sec3}
Tables\,\ref{tavola1} to \ref{tavola3} report the \textcolor{black}{maximal and minimal} values of the LT shifts per orbit of $I$, $\Omega$ and $\omega$ of S4716,  arcseconds per revolution $\ton{^{\prime\prime}\,\mathrm{rev}^{-1}}$, along with the corresponding values for the polar angles $i^\bullet$ and $\varepsilon^\bullet$ of the hole's spin axis. The latter ones are assumed as independent variables in \rfrs{dIdt}{dodt}, whose plots are shown in Figure,\ref{figura1}. It turns out that the \textcolor{black}{maximal} values of the LT orbital shifts per revolution of S4716 are $4.6\,^{\prime\prime}\,\mathrm{rev}^{-1}$ $\ton{\left|\Delta I^\mathrm{LT}\right|_\mathrm{max}}$, $14.4\,^{\prime\prime}\,\mathrm{rev}^{-1}$ $\ton{\left|\Delta\Omega^\mathrm{LT}\right|_\mathrm{max}}$, and $16.4\,^{\prime\prime}\,\mathrm{rev}^{-1}$ $\ton{\left|\Delta\omega^\mathrm{LT}\right|_\mathrm{max}}$. They do not occur for the same values of $i^\bullet$ and $\varepsilon^\bullet$.

\begin{table}[!htb]
\begin{center}
\begin{threeparttable}
\caption{\textcolor{black}{Maximal and Minimal} Nominal Values, in Arcseconds per Revolution $\ton{^{\prime\prime}\,\mathrm{rev}^{-1}}$, of the LT Shift per Orbit $\Delta I^\mathrm{LT}$ of the Inclination $I$ of S4716 Along with the Corresponding Values, in Degrees, of the SMBH's Spin Axis Polar Angles $0^\circ\leq i^\bullet\leq 180^\circ,\,0^\circ\leq\varepsilon^\bullet\leq 360^\circ$.
}\lb{tavola1}
\begin{tabular*}{\textwidth}{c@{\extracolsep{\fill}}c c c c c c c}
\toprule
%
\multirow{2}{*}{}
 &
 $\Delta I_\mathrm{max}^\mathrm{LT}$
 &
 $i^\bullet_\mathrm{max}$
 &
 $\varepsilon^\bullet_\mathrm{max}$
 &
 $\Delta I_\mathrm{min}^\mathrm{LT}$
 &
 $i^\bullet_\mathrm{min}$
 &
 $\varepsilon^\bullet_\mathrm{min}$ \\

 &
 $\ton{^{\prime\prime}\,\mathrm{rev}^{-1}}$
 &
 $\ton{^\circ}$ & $\ton{^\circ}$
 &
 $\ton{^{\prime\prime}\,\mathrm{rev}^{-1}}$
 &
 $\ton{^\circ}$
 &
 $\ton{^\circ}$\\
\midrule
%
%
S4716 & $4.6$ & $90$ & $151.5$ & $-4.6$ & $90$ & $331.5$ \\
\bottomrule
\end{tabular*}
\begin{tablenotes}
\small
\item \textbf{Note.} For  $J_\bullet=\chi_\mathrm{g}\,M_\bullet^2\,G/c$, the values  $\chi_\mathrm{g}=0.5,\,M_\bullet=4.1\times 10^6\,\mathrm{M}_\odot$ \citep{2020ApJ...899...50P,2022ApJ...933...49P} were adopted. According to \rfr{dIdt}, $\dot I^\mathrm{LT}=0^\circ$ for the spin axis aligned with the direction of the line of sight, i.e. for $i^\bullet_0=0^\circ$ or $i^\bullet_0=180^\circ$, or for $\varepsilon^\bullet_0=\Omega\pm 90^\circ$.
\end{tablenotes}
\end{threeparttable}
\end{center}
\end{table}
\begin{table}[!htb]
\begin{center}
\begin{threeparttable}
\caption{\textcolor{black}{Maximal and Minimal} Values, in Arcseconds per Revolution $\ton{^{\prime\prime}\,\mathrm{rev}^{-1}}$, of the LT Shift per Orbit $\Delta\Omega^\mathrm{LT}$ of the Node $\Omega$ of S4716 Along with the Corresponding Values, in Degrees, of the SMBH's Spin Axis Polar Angles $0^\circ\leq i^\bullet\leq 180^\circ,\,0^\circ\leq\varepsilon^\bullet\leq 360^\circ$.
}\lb{tavola2}
\begin{tabular*}{\textwidth}{c@{\extracolsep{\fill}}c c c c c c c c c}
\toprule
%
\multirow{2}{*}{}
&
$\Delta\Omega_\mathrm{max}^\mathrm{LT}$
&
$i^\bullet_\mathrm{max}$
&
$\varepsilon^\bullet_\mathrm{max}$
&
$\Delta\Omega_\mathrm{min}^\mathrm{LT}$
&
$i^\bullet_\mathrm{min}$
&
$\varepsilon^\bullet_\mathrm{min}$
&
$i^\bullet_0$
&
$\varepsilon^\bullet_0$ \\
&
$\ton{^{\prime\prime}\,\mathrm{rev}^{-1}}$
&
$\ton{^\circ}$
&
$\ton{^\circ}$
&
$\ton{^{\prime\prime}\,\mathrm{rev}^{-1}}$
&
$\ton{^\circ}$
&
$\ton{^\circ}$
&
$\ton{^\circ}$
&
$\ton{^\circ}$
\\
\midrule
%
%
S4716 & $14.4$ & $71.2$ & $61.5$ & $-14.4$ & $108.7$ & $241.5$ & $30.4$ & $186.9$\\

\bottomrule
\end{tabular*}
\begin{tablenotes}
\small
\item \textbf{Note.}  For  $J_\bullet=\chi_\mathrm{g}\,M_\bullet^2\,G/c$, the values $\chi_\mathrm{g}=0.5,\,M_\bullet=4.1\times 10^6\,\mathrm{M}_\odot$ \citep{2020ApJ...899...50P,2022ApJ...933...49P} were adopted. The values $i^\bullet_0,\,\varepsilon^\bullet_0$ yield $\dot \Omega^\mathrm{LT}=0^\circ$.
\end{tablenotes}
\end{threeparttable}
\end{center}
\end{table}
\begin{table}[!htb]
\begin{center}
\begin{threeparttable}
\caption{\textcolor{black}{Maximal and Minimal} Values, in Arcseconds per Revolution $\ton{^{\prime\prime}\,\mathrm{rev}^{-1}}$, of the LT Shift per Orbit $\Delta\omega^\mathrm{LT}$ of the Perinigricon $\omega$ of S4716 Along with the Corresponding Values, in Degrees, of the SMBH's Spin Axis Polar Angles $0^\circ\leq i^\bullet\leq 180^\circ,\,0^\circ\leq\varepsilon^\bullet\leq 360^\circ$.
}\lb{tavola3}
\begin{tabular*}{\textwidth}{c@{\extracolsep{\fill}}c c c c c c c c c}
\toprule
%
\multirow{2}{*}{}
&
$\Delta\omega_\mathrm{max}^\mathrm{LT}$
&
$i^\bullet_\mathrm{max}$
&
$\varepsilon^\bullet_\mathrm{max}$
&
$\Delta\omega_\mathrm{min}^\mathrm{LT}$
&
$i^\bullet_\mathrm{min}$
&
$\varepsilon^\bullet_\mathrm{min}$
&
$i^\bullet_0$
&
$\varepsilon^\bullet_0$ \\
&
$\ton{^{\prime\prime}\,\mathrm{rev}^{-1}}$
&
$\ton{^\circ}$
&
$\ton{^\circ}$
&
$\ton{^{\prime\prime}\,\mathrm{rev}^{-1}}$
&
$\ton{^\circ}$
&
$\ton{^\circ}$
&
$\ton{^\circ}$
&
$\ton{^\circ}$
\\
\midrule
%
%
S4716 & $16.4$ & $37.0$ & $61.5$ & $-16.4$ & $142.9$ & $241.5$ & $58.2$ & $276.2$ \\
\bottomrule
\end{tabular*}
\begin{tablenotes}
\small
\item \textbf{Note.} For  $J_\bullet=\chi_\mathrm{g}\,M_\bullet^2\,G/c$, the values $\chi_\mathrm{g}=0.5,\,M_\bullet=4.1\times 10^6\,\mathrm{M}_\odot$ \citep{2020ApJ...899...50P,2022ApJ...933...49P} were adopted. The values $i^\bullet_0,\,\varepsilon^\bullet_0$ yield $\dot\omega^\mathrm{LT}=0^\circ$.
\end{tablenotes}
\end{threeparttable}
\end{center}
\end{table}
\clearpage
\clearpage
A reasonable guess for the measurability of the LT orbital shifts of S4716 is as follows. \textcolor{black}{Assume} that, after each stellar revolution, a determination of any of the orbital elements $\xi$ among $I,\,\Omega,\,\omega$ is obtained; thus, after $N$ orbital periods, the final uncertainty $\sigma_\xi^N$ in $\xi$ should have been reduced down to
\eqi
\sigma_\xi^N\simeq \rp{\sigma_\xi}{N^{1/2}},
\eqf
while the cumulative LT shift of $\xi$ is
\eqi
\Delta\xi^N_\mathrm{LT} = N\,\Delta\xi_\mathrm{LT}.
\eqf
Thus, in order to have
\eqi
\rp{\sigma_\xi^N}{\Delta\xi^N_\mathrm{LT}} \simeq \rp{\sigma_\xi}{N^{3/2}\,\Delta\xi_\mathrm{LT}}= x < 1,
\eqf
the number of revolutions must amount to
\eqi
N \simeq \ton{\rp{\sigma_\xi}{x\,\Delta\xi_\mathrm{LT}}}^{2/3}.\lb{ENNE2}
\eqf
For, say, a $10\%$ test, corresponding to $x = 0.1$, one has
\eqi
N \simeq \ton{\rp{10\,\sigma_\xi}{\Delta\xi_\mathrm{LT}}}^{2/3}.\lb{ENNE}
\eqf
\begin{table}[!htb]
\begin{center}
\begin{threeparttable}
\caption{Current errors $\sigma_I,\,\sigma_\Omega,\,\sigma_\omega$ in the inclination $I$, the longitude of the ascending node $\Omega$, and the argument of perinigricon $\omega$, in degrees, of S4716 according to Table 2 of \citet{2022ApJ...933...49P}.
}\lb{tavola4}
\begin{tabular*}{\textwidth}{c@{\extracolsep{\fill}}c c c}
\toprule
%
\multirow{2}{*}{}
$\sigma_I$
&
$\sigma_\Omega$
&
$\sigma_\omega$\\
$\ton{^\circ}$
&
$\ton{^\circ}$
&
$\ton{^\circ}$
\\
\midrule
%
%
$2.80$ & $1.54$ & $0.02$\\

\bottomrule
\end{tabular*}
\end{threeparttable}
\end{center}
\end{table}
According to the figure for the current error $\sigma_\omega$ in the argument of perinigricon $\omega$, reported in Table\,\ref{tavola4},
the number of revolutions $N$ required to reduce it down to $\simeq 10\%$ of the corresponding LT shift, as per \rfr{ENNE}, is
\eqi
N\simeq 12,\lb{Nomega}
\eqf
corresponding to about
\eqi
\Delta T = 48\,\mathrm{yr}\lb{Tomega}.
\eqf
It should be remarked that \rfrs{Nomega}{Tomega} were obtained by inserting $\Delta\omega_\mathrm{max}^\mathrm{LT}$, quoted in Table\,\ref{tavola3}, in \rfr{ENNE}. As such, one may have to wait even (much?) more time than \rfr{Tomega} since the actual perinigricon shift may be smaller, depending on the orientation of the BH's spin axis.

For the inclination $I$ and the node $\Omega$, much longer time spans would be needed; indeed, from Tables\,\ref{tavola1} to \ref{tavola2}, reporting just $\Delta I_\mathrm{max}^\mathrm{LT},\,\Delta\Omega_\mathrm{max}^\mathrm{LT}$, and Table\,\ref{tavola4}, it turns out
\eqi
N\simeq 250-780,
\eqf
corresponding to $\simeq 1000-3120\,\mathrm{yr}$.
%
%
%
%
\section{The gravitoelectric shift per orbit of S4716 and its measurability}\lb{sec4}
The gravitoelectric perinigricon advance per orbit is
\eqi
\Delta\omega^\mathrm{GE} = \rp{6\,\uppi\,G\,M_\bullet}{c^2\,a\,\ton{1 - e^2}};\lb{GEo}
\eqf
for S4716, it is as large as
\eqi
\Delta\omega^\mathrm{GE} = 907.5\,^{\prime\prime}\,\mathrm{rev}^{-1}.\lb{gep}
\eqf
By applying \rfr{ENNE2} to \rfr{gep}, it turns out that a $1\%\,\ton{x=0.01}$ measurement of it would need
\eqi
N\simeq 4,
\eqf
corresponding to
\eqi
\Delta T\simeq 16\,\mathrm{yr}.
\eqf

By propagating the uncertainties of the hole's mass and of the orbital parameters of S4716 themselves in \rfr{GEo} induce a major systematic bias on the star's LT shift. Suffice it to say that, according to Table 2 of \citet{2022ApJ...933...49P}, the semimajor axis $a$ of S4716 is currently known to a $\simeq 1\%$ accuracy. It yields
\eqi
\sigma_{\Delta\omega^\mathrm{GE}} \simeq 9.5\,^{\prime\prime}\,\mathrm{rev}^{-1},
\eqf
which corresponds to $\simeq 57\%$ of $\left|\Delta\omega^\mathrm{LT}\right|_\mathrm{max}$.
\textcolor{black}{This is a major issue since,  even if an observation program long enough for the gravitomagnetic
component of the perinigricon motion of S4716 to show itself were undertaken, the observations would render only the sum of the LT and the GR shifts.
A possible way out would be monitoring the perinigricon precessions of more stars and linearly combining them in such a way to disentangle the relativistic contributions to them, as proposed for the first time, in a different scenario, by \citet{1990grg..conf..313S}. See also Section 5 of \citet{2019EPJC...79..816I},  and \citet{2008ApJ...674L..25W}.
}
\section{The quadrupole-induced orbital shifts}\lb{sec5}
A spinning, oblate body, endowed with a quadrupole mass moment $Q_2$ arbitrarily oriented in space, induces the following orbital precessions \citep{1975PhRvD..12..329B,1999ApJ...514..388W,2017EPJC...77..439I}
\begin{align}
\dot I^{Q_2} \lb{IQ2} & = \rp{3\,\nk\,Q_2}{2\,a^2\,\ton{1-e^2}^2\,M}\,\ton{\bds{\hat{J}}\bds\cdot\bds{\hat{l}}}\ton{\bds{\hat{J}}\bds\cdot\bds{\hat{h}}}, \\ \nonumber \\
\dot \Omega^{Q_2} \lb{OQ2} & = \rp{3\,\nk\,Q_2}{2\,a^2\,\ton{1-e^2}^2\,M}\,\csc I\,\ton{\bds{\hat{J}}\bds\cdot\bds{\hat{m}}}\ton{\bds{\hat{J}}\bds\cdot\bds{\hat{h}}}, \\ \nonumber \\
\dot \omega^{Q_2} \lb{oQ2} & = -\rp{3\,\nk\,Q_2}{4\,a^2\,\ton{1-e^2}^2\,M}\,\grf{
2 - 3\,\qua{
\ton{\bds{\hat{J}}\bds\cdot\bds{\hat{l}}}^2 + \ton{\bds{\hat{J}}\bds\cdot\bds{\hat{m}}}^2
} + 2\,\cot I\,\ton{\bds{\hat{J}}\bds\cdot\bds{\hat{m}}}\ton{\bds{\hat{J}}\bds\cdot\bds{\hat{h}}}
}
\end{align}
where $\nk\doteq\sqrt{G\,M/a^3}$ is the Keplerian mean motion. Here, $Q_2$ is negative definite, and has the dimensions of a mass times a squared length; it can be expressed as
\eqi
Q_2 = - J_2\,M\,R^2
\eqf
in terms of the body's equatorial radius $R$ and of the positive definite, dimensionless first even zonal harmonic $J_2$ of the multipolar expansion of its non-spherical gravitational potential.

According to the no-hair theorems \citep{1967PhRv..164.1776I,1968CMaPh...8..245I,1971PhRvL..26..331C}, the quadrupole mass moment $Q_2^\bullet$ of a spinning black hole  is \citep{1970JMP....11.2580G,1974JMP....15...46H}
\eqi
Q_2^\bullet = -\rp{J_\bullet^2}{c^2 M_\bullet}.
\eqf

In the case of S4716, the effect of the quadrupole of the SMBH in Sgr A$^\ast$ is negligible; according to \rfrs{IQ2}{oQ2}, the $Q_2^\bullet$-induced shifts per orbit amount to a maximum of $\simeq 0.01-0.05\,^{\prime\prime}\,\mathrm{rev}^{-1}$.
\section{Summary and conclusions}\lb{sec6}
The star S4716 star is, at the time of writing, the member of the stellar cluster around the supermassive black hole in Sgr A$^\ast$ exhibiting the shortest orbital period.

Its LT orbital precessions, affecting $I$, $\Omega$ and $\omega$, depend, among other things, also on the orientation of the hole's spin axis in space, which is poorly constrained; for certain spin configurations, differing from one orbital element to another, they vanish. By treating the spin's polar angles determining its position in space as free parameters, it turns out that the \textcolor{black}{maximal} values of the star's LT shifts per orbit are of the order of $\simeq 5-16\,^{\prime\prime}\,\mathrm{rev}^{-1}$. Given the current error in measuring the perinigricon, which is the most accurately determined orbital element of S4716 among all those impacted by the hole's gravitomagnetic field, it would took about 48 yr to bring it down to $10\%$ of the cumulative LT shift. Much longer temporal intervals would be required for measuring the LT shifts of $I$ and $\Omega$ to the same level of accuracy.

A $\simeq 1\%$ measurement of the larger Schwarzschild-like perinigricon shift would be possible after just 16 yr. On the other hand, the uncertainties in the physical and orbital parameters entering its expression would induce a systematic error amounting to a significative part of the LT effect for $\omega$.

The impact of the hole's quadrupole is negligible since its shifts per orbit are as little as $\simeq 0.01-0.05\,^{\prime\prime}\,\mathrm{rev}^{-1}$.

\bibliography{Sstarsbib}{}

\begin{thebibliography}{45}
\expandafter\ifx\csname natexlab\endcsname\relax\def\natexlab#1{#1}\fi

\bibitem[{{Ali} {et~al}\mbox{.}(2020){Ali}, {Paul}, {Eckart}, {Parsa},
  {Zajacek}, {Pei{\ss}ker}, {Subroweit}, {Valencia-S.}, {Thomkins}, \&
  {Witzel}}]{2020ApJ...896..100A}
{Ali} B. {et~al.}, 2020, \apj, 896, 100

\bibitem[{{Alush} \& {Stone}(2022)}]{2022PhRvD.106l3023A}
{Alush} Y., {Stone} N.~C., 2022, \prd, 106, 123023

\bibitem[{{Ang{\'e}lil}, {Saha} \& {Merritt}(2010){Ang{\'e}lil}, {Saha}, \&
  {Merritt}}]{2010ApJ...720.1303A}
{Ang{\'e}lil} R., {Saha} P., {Merritt} D., 2010, \apj, 720, 1303

\bibitem[{{Barker} \& {O'Connell}(1975)}]{1975PhRvD..12..329B}
{Barker} B.~M., {O'Connell} R.~F., 1975, PhRvD, 12, 329

\bibitem[{{Carter}(1971)}]{1971PhRvL..26..331C}
{Carter} B., 1971, \prl, 26, 331

\bibitem[{{Chandrasekhar}(1983)}]{1983mtbh.book.....C}
{Chandrasekhar} S., 1983, {The mathematical theory of black holes}. Clarendon
  Press, Oxford

\bibitem[{{Damour} \& {Sch\"{a}fer}(1988)}]{1988NCimB.101..127D}
{Damour} T., {Sch\"{a}fer} G., 1988, NCimB, 101, 127

\bibitem[{{Damour} \& {Taylor}(1992)}]{1992PhRvD..45.1840D}
{Damour} T., {Taylor} J.~H., 1992, PhRvD, 45, 1840

\bibitem[{{Einstein}(1915)}]{Ein15}
{Einstein} A., 1915, Sitzungsber. Kgl. Preuss. Akad. Wiss., 47, 831

\bibitem[{{Fragione} \& {Loeb}(2020)}]{2020arXiv200811734F}
{Fragione} G., {Loeb} A., 2020, ApJL, 901, L32

\bibitem[{{Fragione} \& {Loeb}(2022)}]{2022ApJ...932L..17F}
{Fragione} G., {Loeb} A., 2022, \apjl, 932, L17

\bibitem[{{Genzel}, {Eisenhauer} \& {Gillessen}(2010){Genzel}, {Eisenhauer}, \&
  {Gillessen}}]{2010RvMP...82.3121G}
{Genzel} R., {Eisenhauer} F., {Gillessen} S., 2010, Rev. Mod. Phys., 82, 3121

\bibitem[{{Geroch}(1970)}]{1970JMP....11.2580G}
{Geroch} R., 1970, JMP, 11, 2580

\bibitem[{{Ghez} {et~al}\mbox{.}(2008){Ghez}, {Salim}, {Weinberg}, {Lu}, {Do},
  {Dunn}, {Matthews}, {Morris}, {Yelda}, {Becklin}, {Kremenek},
  {Milosavljevic}, \& {Naiman}}]{2008ApJ...689.1044G}
{Ghez} A.~M. {et~al.}, 2008, \apj, 689, 1044

\bibitem[{{Gillessen} {et~al}\mbox{.}(2008){Gillessen}, {Genzel}, {Eisenhauer},
  {Ott}, {Trippe}, \& {Martins}}]{2008IAUS..248..466G}
{Gillessen} S., {Genzel} R., {Eisenhauer} F., {Ott} T., {Trippe} S., {Martins}
  F., 2008, in IAU Symposium, Vol. 248, A Giant Step: from Milli- to
  Micro-arcsecond Astrometry, {Jin} W.~J., {Platais} I., {Perryman} M.~A.~C.,
  eds., pp. 466--469

\bibitem[{{Gravity Collaboration} {et~al}\mbox{.}(2020){Gravity Collaboration},
  {Abuter}, {Amorim}, {Baub{\"o}ck}, {Berger}, {Bonnet}, {Brand ner},
  {Cardoso}, {Cl{\'e}net}, {de Zeeuw}, {Dexter}, {Eckart}, {Eisenhauer},
  {F{\"o}rster Schreiber}, {Garcia}, {Gao}, {Gendron}, {Genzel}, {Gillessen},
  {Habibi}, {Haubois}, {Henning}, {Hippler}, {Horrobin}, {Jim{\'e}nez-Rosales},
  {Jochum}, {Jocou}, {Kaufer}, {Kervella}, {Lacour}, {Lapeyr{\`e}re}, {Le
  Bouquin}, {L{\'e}na}, {Nowak}, {Ott}, {Paumard}, {Perraut}, {Perrin},
  {Pfuhl}, {Rodr{\'\i}guez-Coira}, {Shangguan}, {Scheithauer}, {Stadler},
  {Straub}, {Straubmeier}, {Sturm}, {Tacconi}, {Vincent}, {von Fellenberg},
  {Waisberg}, {Widmann}, {Wieprecht}, {Wiezorrek}, {Woillez}, {Yazici}, \&
  {Zins}}]{2020A&A...636L...5G}
{Gravity Collaboration} {et~al.}, 2020, \aap, 636, L5

\bibitem[{{Han}(2014)}]{2014RAA....14.1415H}
{Han} W.-B., 2014, RAA, 14, 1415

\bibitem[{{Hansen}(1974)}]{1974JMP....15...46H}
{Hansen} R.~O., 1974, JMP, 15, 46

\bibitem[{{Iorio}(2011)}]{2011MNRAS.411..453I}
{Iorio} L., 2011, \mnras, 411, 453

\bibitem[{{Iorio}(2017)}]{2017EPJC...77..439I}
{Iorio} L., 2017, EPJC, 77, 439

\bibitem[{{Iorio}(2019)}]{2019EPJC...79..816I}
{Iorio} L., 2019, European Physical Journal C, 79, 816

\bibitem[{{Israel}(1967)}]{1967PhRv..164.1776I}
{Israel} W., 1967, Phys. Rev., 164, 1776

\bibitem[{{Israel}(1968)}]{1968CMaPh...8..245I}
{Israel} W., 1968, Commun. Math. Phys., 8, 245

\bibitem[{{Jaroszynski}(1998)}]{1998AcA....48..653J}
{Jaroszynski} M., 1998, AcA, 48, 653

\bibitem[{{Kerr}(1963)}]{1963PhRvL..11..237K}
{Kerr} R.~P., 1963, \prl, 11, 237

\bibitem[{{Kraniotis}(2007)}]{2007CQGra..24.1775K}
{Kraniotis} G.~V., 2007, CQGra, 24, 1775

\bibitem[{{Levin} \& {Beloborodov}(2003)}]{2003ApJ...590L..33L}
{Levin} Y., {Beloborodov} A.~M., 2003, ApJL, 590, L33

\bibitem[{{Lynden-Bell} \& {Rees}(1971)}]{1971MNRAS.152..461L}
{Lynden-Bell} D., {Rees} M.~J., 1971, \mnras, 152, 461

\bibitem[{{Merritt}(2013)}]{2013degn.book.....M}
{Merritt} D., 2013, {Dynamics and Evolution of Galactic Nuclei}. Princeton
  University Press, Princeton

\bibitem[{{Merritt} {et~al}\mbox{.}(2010){Merritt}, {Alexander}, {Mikkola}, \&
  {Will}}]{2010PhRvD..81f2002M}
{Merritt} D., {Alexander} T., {Mikkola} S., {Will} C.~M., 2010, PhRvD, 81,
  062002

\bibitem[{{Meyer} {et~al}\mbox{.}(2007){Meyer}, {Sch{\"o}del}, {Eckart},
  {Duschl}, {Karas}, \& {Dov{\v{c}}iak}}]{2007A&A...473..707M}
{Meyer} L., {Sch{\"o}del} R., {Eckart} A., {Duschl} W.~J., {Karas} V.,
  {Dov{\v{c}}iak} M., 2007, \aap, 473, 707

\bibitem[{{Nobili} \& {Will}(1986)}]{1986Natur.320...39N}
{Nobili} A.~M., {Will} C.~M., 1986, Nature, 320, 39

\bibitem[{{Pei{\ss}ker} {et~al}\mbox{.}(2020){Pei{\ss}ker}, {Eckart},
  {Zaja{\v{c}}ek}, {Ali}, \& {Parsa}}]{2020ApJ...899...50P}
{Pei{\ss}ker} F., {Eckart} A., {Zaja{\v{c}}ek} M., {Ali} B., {Parsa} M., 2020,
  \apj, 899, 50

\bibitem[{{Pei{\ss}ker} {et~al}\mbox{.}(2022){Pei{\ss}ker}, {Eckart},
  {Zaja{\v{c}}ek}, \& {Britzen}}]{2022ApJ...933...49P}
{Pei{\ss}ker} F., {Eckart} A., {Zaja{\v{c}}ek} M., {Britzen} S., 2022, \apj,
  933, 49

\bibitem[{{Penrose}(2002)}]{2002GReGr..34.1141P}
{Penrose} R., 2002, Gen. Relativ. Gravit., 7, 1141

\bibitem[{{Preto} \& {Saha}(2009)}]{2009ApJ...703.1743P}
{Preto} M., {Saha} P., 2009, \apj, 703, 1743

\bibitem[{{Psaltis}(2019)}]{2019GReGr..51..137P}
{Psaltis} D., 2019, GReGr, 51, 137

\bibitem[{{Psaltis}, {Wex} \& {Kramer}(2016){Psaltis}, {Wex}, \&
  {Kramer}}]{2016ApJ...818..121P}
{Psaltis} D., {Wex} N., {Kramer} M., 2016, \apj, 818, 121

\bibitem[{{Shapiro}(1990)}]{1990grg..conf..313S}
{Shapiro} I.~I., 1990, in General Relativity and Gravitation, 1989, {Ashby} N.,
  {Bartlett} D.~F., {Wyss} W., eds., Cambridge University Press, Cambridge, pp.
  313--330

\bibitem[{{Waisberg} {et~al}\mbox{.}(2018){Waisberg}, {Dexter}, {Gillessen},
  {Pfuhl}, {Eisenhauer}, {Plewa}, {Baub{\"o}ck}, {Jimenez-Rosales}, {Habibi},
  {Ott}, {von Fellenberg}, {Gao}, {Widmann}, \& {Genzel}}]{2018MNRAS.476.3600W}
{Waisberg} I. {et~al.}, 2018, \mnras, 476, 3600

\bibitem[{{Wex} \& {Kopeikin}(1999)}]{1999ApJ...514..388W}
{Wex} N., {Kopeikin} S.~M., 1999, \apj, 514, 388

\bibitem[{{Will}(2008)}]{2008ApJ...674L..25W}
{Will} C.~M., 2008, \apjl, 674, L25

\bibitem[{{Yu}, {Zhang} \& {Lu}(2016){Yu}, {Zhang}, \&
  {Lu}}]{2016ApJ...827..114Y}
{Yu} Q., {Zhang} F., {Lu} Y., 2016, ApJ, 827, 114

\bibitem[{{Zhang} \& {Iorio}(2017)}]{2017ApJ...834..198Z}
{Zhang} F., {Iorio} L., 2017, \apj, 834, 198

\bibitem[{{Zhang}, {Lu} \& {Yu}(2015){Zhang}, {Lu}, \&
  {Yu}}]{2015ApJ...809..127Z}
{Zhang} F., {Lu} Y., {Yu} Q., 2015, \apj, 809, 127

\end{thebibliography}

\end{document}